\UseRawInputEncoding
\documentclass{raa}            

\usepackage{graphicx,times}
\usepackage{natbib}
\usepackage{amssymb,amsmath}
\bibpunct{(}{)}{;}{a}{}{,}

\usepackage[pagebackref=true]{hyperref}
\hypersetup{colorlinks = true, linkcolor = green, anchorcolor = red, citecolor = blue, filecolor = red, pagecolor = red, urlcolor = red}

\begin{document}

  \title{Sympathetic solar eruption on 2024 February 9}

   \volnopage{Vol.0 (2024) No.0, 000--000}
   \setcounter{page}{1} 

   \author{Shu-Yue Li
      \inst{1,2}
   \and Qing-Min Zhang
      \inst{2}
   \and Bei-Li Ying
      \inst{2}
   \and Li Feng
      \inst{2}
  \and Ying-Na Su
      \inst{2}
  \and Mu-Sheng Lin
      \inst{2,3}
  \and Yan-Jie Zhang
      \inst{2}
   }

   \institute{School of Science, Nanjing University of Posts and Telecommunications, Nanjing 210023, China; {\it zhangqm@pmo.ac.cn}\\
        \and
             Key Laboratory of Dark Matter and Space Astronomy, Purple Mountain Observatory, Nanjing 210023, China\\
        \and
             School of Astronomy and Space Science, University of Science and Technology of China, Hefei 230026, China\\
\vs\no
   {\small Received 2024 month day; accepted 202x month day}}

\abstract{In this paper, we perform a follow-up investigation of the solar eruption originating from active region (AR) 13575 on 2024 February 9.
The primary eruption of a hot channel (HC) generates an X3.4 class flare, a full-halo coronal mass ejection (CME), and an extreme-ultraviolet (EUV) wave.
Interaction between the wave and a quiescent prominence (QP) leads to a large-amplitude, transverse oscillation of QP.
After the transverse oscillation, QP loses equilibrium and rises up.
The ascending motion of the prominence is coherently detected and tracked up to $\sim$1.68 $R_{\odot}$ by the Solar UltraViolet Imager (SUVI) onboard the GOES-16 spacecraft
and up to $\sim$2.2 $R_{\odot}$ by the Solar Corona Imager (SCI\_UV) of the Lyman-alpha Solar Telescope (LST) onboard the ASO-S spacecraft.
The velocity increases linearly from 12.3 to 68.5 km s$^{-1}$ at 18:30 UT.
The sympathetic eruption of QP drives the second CME with a typical three-part structure. 
The bright core comes from the eruptive prominence, which could be further observed up to $\sim$3.3 $R_{\odot}$ 
by the Large Angle Spectroscopic Coronagraph (LASCO) onboard the SOHO mission.
The leading edge of the second CME accelerates continuously from $\sim$120 to $\sim$277 km s$^{-1}$.
The EUV wave plays an important role in linking the primary eruption with the sympathetic eruption.
\keywords{Sun: flares --- Sun: filaments, prominences --- Sun: coronal mass ejections (CMEs)}
}

   \authorrunning{Zhang et al.}
   \titlerunning{Sympathetic solar eruption}

   \maketitle

\section{Introduction} \label{intro}
Filaments are widespread in the solar corona (\citealt{eng98,mac10,par14}). 
They consist of very dynamic and dark threads observed in H$\alpha$ line center (\citealt{mar98,lin11}).
Above the limb, filaments are called prominences, which show bright features due to their strong emissions 
in H$\alpha$ (6562.8 {\AA}), Ca\,{\sc ii} H line (3968 {\AA}), H\,{\sc i} Lyman-$\alpha$ (1216 {\AA}), and He\,{\sc ii} (304 {\AA}) wavebands (\citealt{ber08,zhou23,qiu24,xue24}).
After being disturbed, the filament threads tend to deviate from their equilibrium positions and oscillate for a few cycles before calming down (\citealt{oli02,tri09,arr18}).

According to the velocity amplitude, filament oscillations are divided into small-amplitude ($\leq$10 km s$^{-1}$) oscillations (SAOs) and large-amplitude ($\geq$20 km s$^{-1}$) oscillations (LAOs).
SAOs are frequently observed in quiescent prominences (\citealt{oka07,ning09,li18,wang24}).
LAOs are occasionally detected in active region (AR) filaments as well as quiescent filaments (\citealt{shen14a,luna18,luna24}).
Based on the direction of oscillations, LAOs are further divided into longitudinal and transverse oscillations.
The filament materials move along the threads during longitudinal oscillations (\citealt{jing03,vrs07,zqm12,zqm17,zheng17,ni22}).
On the contrary, the filament threads swing back and forth perpendicular to the spine during transverse oscillations (\citealt{iso06,her11,dai23,zyj24}).
\citet{shen14b} discovered simultaneous transverse oscillations of one filament and longitudinal oscillations of another one induced by a single shock wave.
Longitudinal and transverse oscillations before eruptions have been observed (\citealt{iso06,bi14,zqm20,dai21,ni22}),
which makes LAOs a credible precursor for filament eruptions considering that LAOs may stand for a transition phase between the initial equilibrium and the eventual eruption (\citealt{chen08,zqm12,fan20}).

Filament eruptions are intimately related with solar flares (\citealt{fle11}) and coronal mass ejections (CMEs; \citealt{for06,chen11}) in the standard flare model, 
namely the CSHKP model (\citealt{car64,stu66,hir74,kopp76}).
Owing to the magnetic connectivity between two or three filaments,
the eruption of a primary filament may considerably affect the environment of another filament and lead to a second eruption, which is considered as a sympathetic eruption.
The most commonplace condition for a sympathetic eruption is that multiple filaments are suppressed by a common, overlying magnetic system.
Successful eruption of the first filament may significantly reduce the constraint on the adjacent filaments and trigger off a second 
or even a chain of eruptions (\citealt{ste04,cheng05,shen12,yang12,cx13,lyn13,jos16,wang16,wang18,dac18,hou20,song20,zhou21}).
\citet{tor11} performed a three-dimensional (3D) magnetohydrodynamics (MHD) simulation to explore possible magnetic trigger mechanisms for sympathetic eruptions.
In their model, two magnetic flux ropes (MFRs) are located within a pseudo-streamer and the third one is located nearby.
The eruption of the third rope gives rise to consecutive eruptions of the two MFRs, which could excellently explain the observed twin filament eruptions and CMEs on 2010 August 1 (\citealt{tit12}).
\citet{shen12} studied two sympathetic filament eruptions including a partial and full MFR eruptions in a quadrupolar magnetic configuration on 2011 May 12.
A schematic model is proposed to illustrate the whole process. Breakout magnetic reconnection occurs after the first filament rises up, which finally evolves into a CME.
The adjacent filament undergoes a partial eruption, with the top part evolving into a blob.

Impulsive eruptions are likely to drive coronal extreme-ultraviolet (EUV) waves (\citealt{tho98,chen02,pats09}) 
or chromospheric Moreton waves (\citealt{mor60,eto02}), which propagate far away at speeds of hundreds of to $\sim$1000 km s$^{-1}$.
Interactions between these global waves and remote filaments may result in LAOs (\citealt{gil08,her11,asai12,liu13,dai23,zyj24}) or direct eruptions.
\citet{jiang11} investigated three successive filament eruptions from different locations on 2003 November 19. The first eruption originates from AR 10501 and generates a CME. 
The CME-related coronal dimmings propagate outward and interact with two quiescent filaments, leading to the second and third eruptions and related flares.
It is concluded that the dimming process in the first eruption results in weakening and partial removal of the large-scale, overlying magnetic fields on the two remote filaments, 
which facilitates sympathetic eruptions.
\citet{dai21} studied the sympathetic eruption of a very long quiescent filament excited by the eruption of a nearby smaller filament on 2015 April 28. The two parallel filaments are $\sim$250 Mm apart.
Prior to the sympathetic eruption, the huge filament undergoes both longitudinal and transverse oscillations, which is accompanied with continuous mass drainage at speeds of 35$-$85 km s$^{-1}$.
The combination of LAOs and mass drainage indicates that the filament is losing equilibrium gradually.
Till now, a complete process of LAOs and the subsequent sympathetic eruption of a prominence excited by EUV waves has not been reported.
Using multiwavelength observations of the Atmospheric Imaging Assembly (AIA; \citealt{lem12}) onboard the Solar Dynamics Observatory (SDO; \citealt{pes12}),
\citet{zqm24a} (hereafter Paper I) studied two successive EUV waves and the induced transverse oscillation of a quiescent prominence (QP) on 2024 February 9.
The EUV waves are separately driven by a fast halo CME \footnote{https://cdaw.gsfc.nasa.gov/CME\_list/UNIVERSAL\_ver2/2024\_02/univ2024\_02.html}
(hereafter CME1) as a result of a hot channel (HC) eruption and by a fast coronal jet originating from AR 13575. 
QP is close to the solar South pole and is more than 380 Mm away from the flare site.

In this paper, using EUV observations of the Solar UltraViolet Imager (SUVI; \citealt{sea18,tad19,dar22}) 
onboard the Geostationary Operational Environmental Satellite (GOES-16) 
and UV (121.6$\pm$10 nm) observations of the Solar Corona Imager (SCI) of the Lyman-alpha (Ly$\alpha$) Solar Telescope (LST; \citealt{feng19,li19,chen24}) 
onboard the Advanced Space-based Solar Observatory (ASO-S; \citealt{gan19,gan23}), 
we reanalyze the event, focusing on the sympathetic eruption of QP and the related CME (hereafter CME2).
Characteristics of these instruments are summarized in Table~\ref{tab-1}.
In Section~\ref{data}, we briefly describe the data analysis of these instruments.
In Section~\ref{res}, we first show the primary eruption, which leads to an X3.4 flare, CME1, and an EUV wave.
Then, we show the sympathetic eruption of QP, which leads to CME2 without a flare.
Discussions and a brief summary are arranged in Section~\ref{dis} and Section~\ref{sum}, respectively.

\begin{table}
	\centering
	\caption{Wavelengths, pixel sizes, cadences, and field of views (FOVs) of the instruments on 2024 February 9.}
	\label{tab-1}
	\begin{tabular}{ccccc}
		\hline
		Instrument & $\lambda$ & Pixel size & Cadence & FOV \\
		                  & ({\AA})     & (arcsec)    &  (s)          & ($R_{\odot}$) \\
		\hline
		SDO/AIA & 211 & 0.6 & 12 & $\sim$1.3 \\
		GOES-16/SUVI & 304 & 2.5 & $\sim$120 & $\sim$1.6 \\
		GOES-16/SUVI & 171 & 2.5 & 240 & $\sim$1.6 \\
		GOES-16          & 1$-$8 & -- & 1 & -- \\
		ASO-S/SCI\_UV & 1216 & 2.15 & $\sim$60 & 1.1$-$2.5 \\
		SOHO/LASCO-C2 & WL & 11.4 & 720 & 2$-$6 \\
		SOHO/LASCO-C3 & WL & 56.0 & 720 & 4$-$30 \\
		STA/COR2 & WL & 15 & 900 & 2.5$-$15 \\
		\hline
	\end{tabular}
\end{table}

\section{Data analysis} \label{data}
SDO/AIA takes full-disk images in 7 EUV (94, 131, 171, 193, 211, 304, and 335 {\AA}) and 2 UV (1600 and 1700 {\AA}) wavelengths.
The level\_1 data are calibrated using the standard program aia\_prep.pro in the \textit{Solar Software} package.
The full-disk GOES-16/SUVI images in 171 and 304 {\AA} are rotated and slightly shifted to align with the AIA images \citep{zqm24b}.
The Ly$\alpha$ images from ASO-S/SCI\_UV are primarily processed through the correction of flat fields and dark currents, along with the normalization of exposure times. 
To suppress stray light signals, we subtract the daily-minimum background from the images. 
Following this, the images are rotated, scaled, and aligned by overlapping structures in the field of view (FOV) with the AIA 304 {\AA} images.
The high cadence and large FOV of ASO-S/SCI\_UV provide a great opportunity to track the erupting prominence from $\sim$1.1 to $\sim$2.2 $R_{\odot}$ in Ly$\alpha$ wavelength.
The white-light (WL) images of CMEs are obtained from the C2 and C3 coronagraphs of the Large Angle Spectroscopic Coronagraph \citep[LASCO;][]{bru95} 
onboard the Solar and Heliospheric Observatory (SOHO) mission 
as well as the COR2 coronagraph on board the ahead Solar TErrestrial RElations Observatory \citep[STEREO;][]{kai08}.

\section{Results} \label{res}
\subsection{Primary eruption} \label{prim} 
\begin{figure}
\centering
\includegraphics[width=0.8\columnwidth]{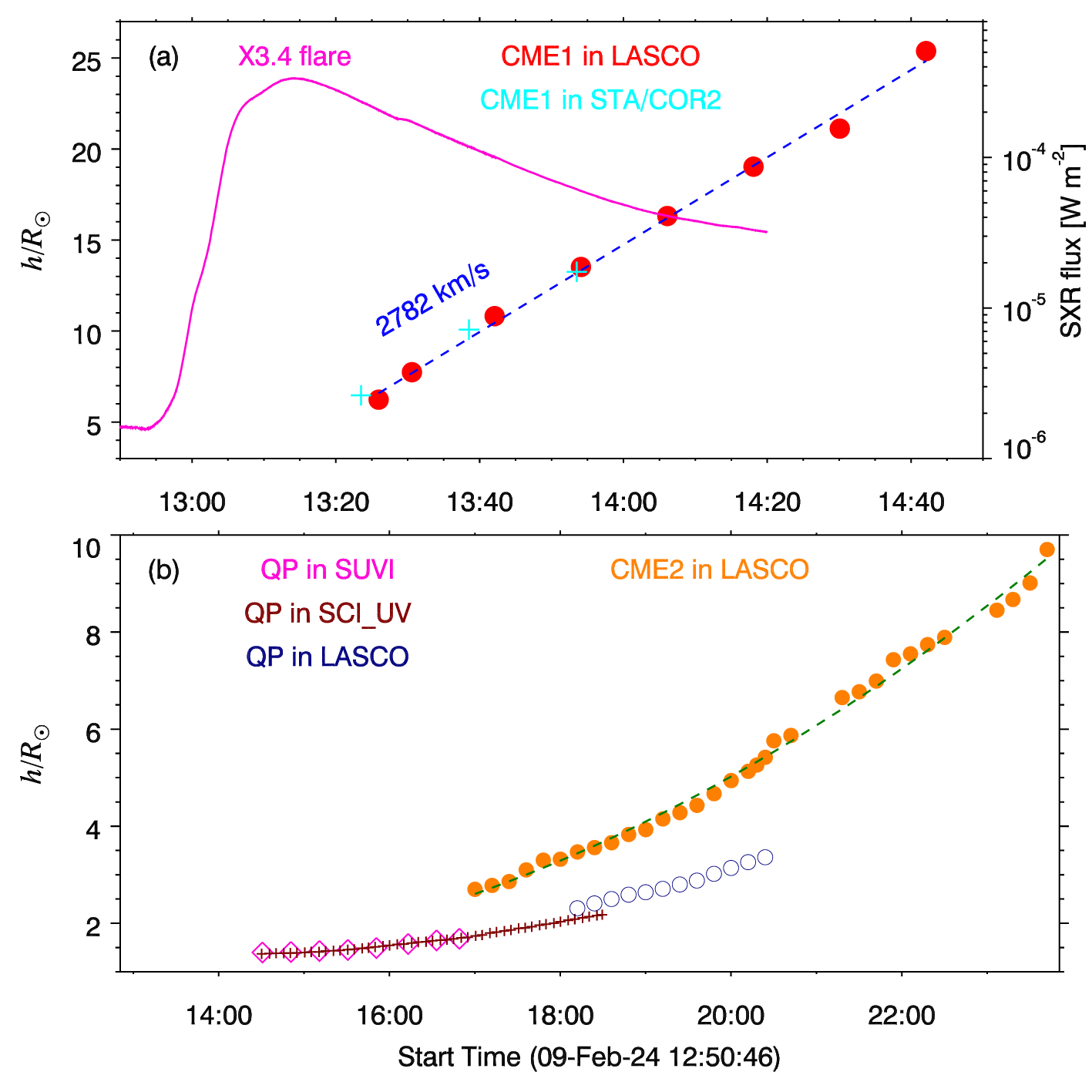}
    \caption{(a) Soft X-ray (SXR) light curve of the X3.4 class flare in 1$-$8 {\AA} (magenta line) and height evolutions of CME1 in LASCO (red circles) and STA/COR2 (cyan pluses).
    A linear fitting is performed with a velocity of $\sim$2782 km s$^{-1}$ for CME1 in the FOV of LASCO.
    (b) Height evolutions of QP in SUVI (magenta diamonds), SCI\_UV (brown pluses), and LASCO (blue circles).
    Height evolution of CME2 is plotted with orange circles and fitted with a quadratic function (green dashed line).}
    \label{fig1}
\end{figure}

As is shown in Fig. 3 and Fig. 5 of Paper I, the HC is exclusively observed in 131 and 94 {\AA} of SDO/AIA. 
It undergoes a slow rise phase and a fast rise phase, the latter of which starts from 12:53:53 UT and lasts for $\sim$10 minutes before escaping the FOV of AIA.
In Figure~\ref{fig1}(a), soft X-ray (SXR) light curve of the flare in 1$-$8 {\AA} recorded by the GOES-16 spacecraft is plotted with a magenta line.
The SXR emission increases from 12:53 UT and peaks at 13:14 UT before declining gradually.

\begin{figure}
\centering
	\includegraphics[width=0.9\columnwidth]{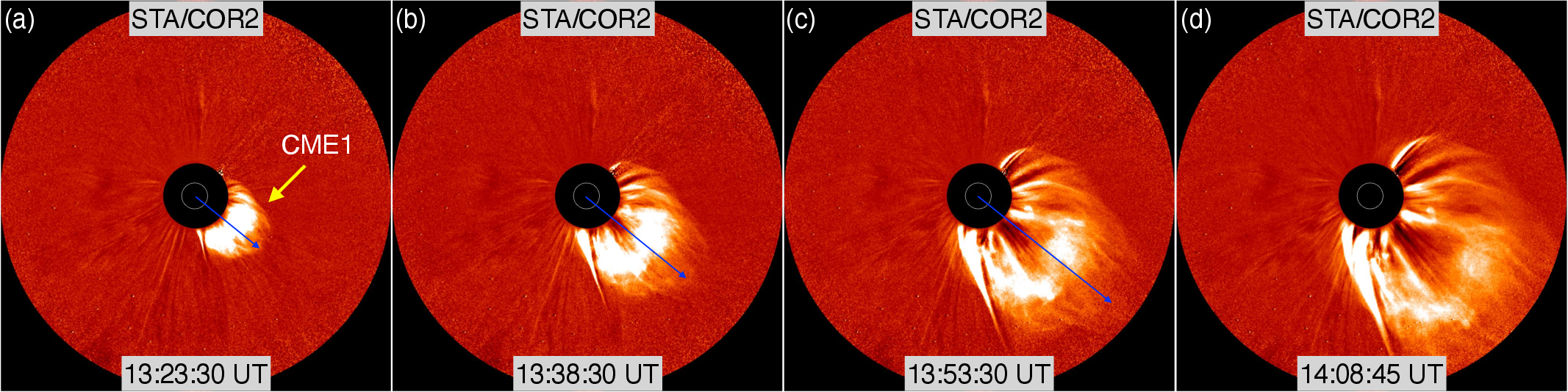}
    \caption{CME1 observed by STA/COR2 during 13:23$-$14:09 UT. The blue arrows indicate the propagation direction and heliocentric distances of the CME leading edge.}
    \label{fig2}
\end{figure}

Figure~\ref{fig2} shows four running-difference images observed by STA/COR2, which had a separation angle of 7.6$^{\circ}$ with the Sun-Earth connection on 2024 Februrary 9.
In panel (a), the yellow arrow points to CME1 when it first appears at 13:23:30 UT.
CME1 propagates in the southwest direction and expands rapidly to form a full-halo CME (panels (b)-(d)).
In panels (a)-(c), the blue arrows indicate the propagation direction and heliocentric distances of the CME leading edge, which are plotted with cyan pluses in Figure~\ref{fig1}(a). 

As is shown in the bottom panels of Fig. 3 in Paper I, CME1 was also observed by SOHO/LASCO.  
The height variation of CME1 with time in the FOV of SOHO/LASCO is plotted with red circles in Figure~\ref{fig1}(a).
A linear fitting is performed between 13:25 UT and 14:45 UT, which is displayed with a blue dashed line. The apparent speed of CME1 reaches $\sim$2782 km s$^{-1}$ in the plane of the sky.

\begin{figure}
\centering
	\includegraphics[width=0.9\columnwidth]{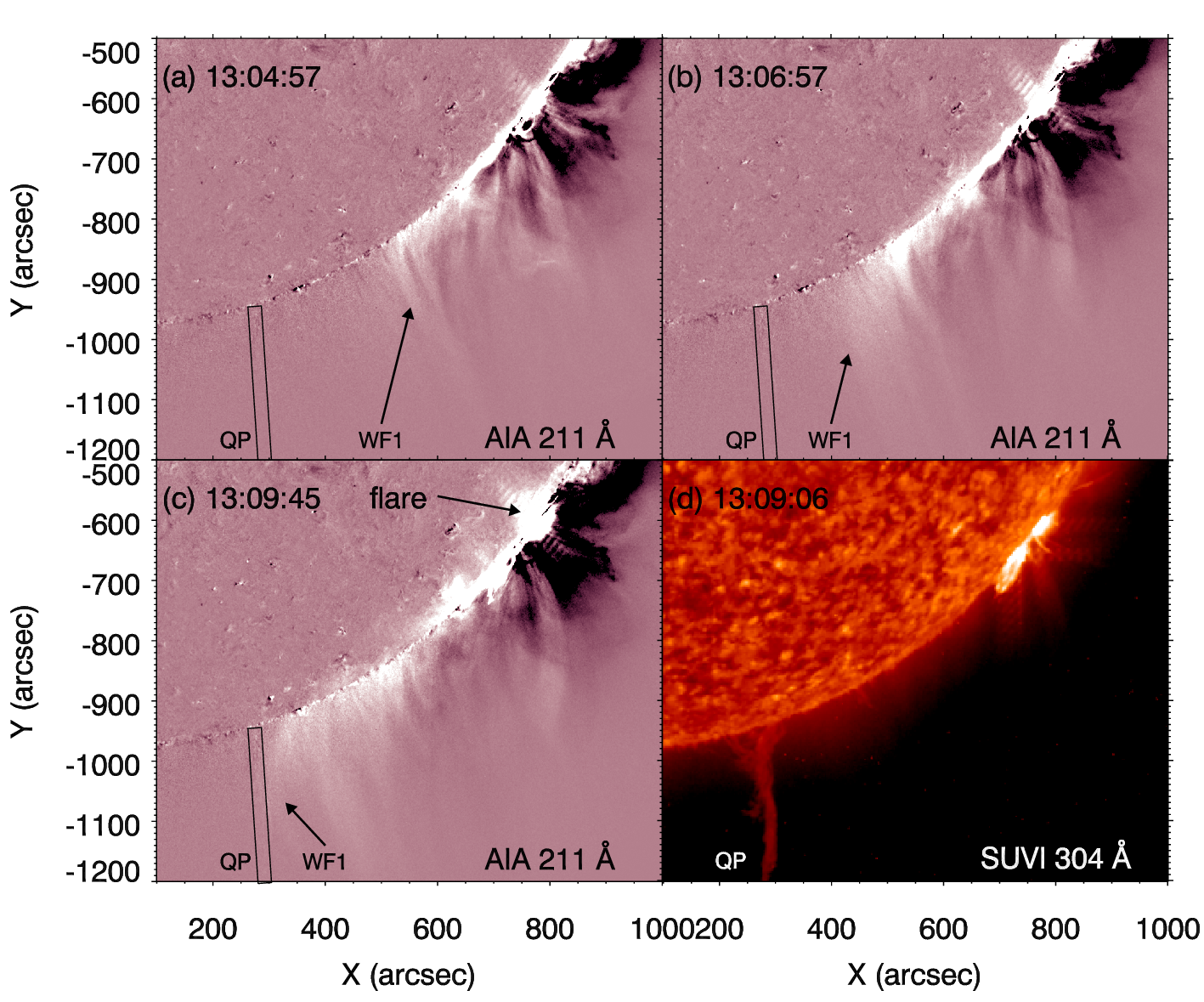}
    \caption{(a-c) AIA 211 {\AA} base-difference images, showing the flare and EUV wave front (WF1). The position of QP is denoted by black rectangles.
    (d) SUVI 304 {\AA} image at 13:09:06 UT, showing the columnar QP near the south polar region.}
    \label{fig3}
\end{figure}

In Figure~\ref{fig3}, panels (a-c) show base-difference images in AIA 211 {\AA} during 13:05$-$13:09 UT (see also Fig. 6 in Paper I). 
QP is denoted by black rectangles. Panel (d) shows the GOES-16/SUVI 304 {\AA} image at 13:09:06 UT, featuring the columnar QP near the south polar region.
The quick expansion of CME1 generates an EUV wave front (WF1), which is indicated by black arrows.
WF1 propagates in the southeast direction at a speed of $\sim$835 km s$^{-1}$ and arrives at the prominence at $\sim$13:09 UT (panel (c)).

\begin{figure}
\centering
	\includegraphics[width=0.9\columnwidth]{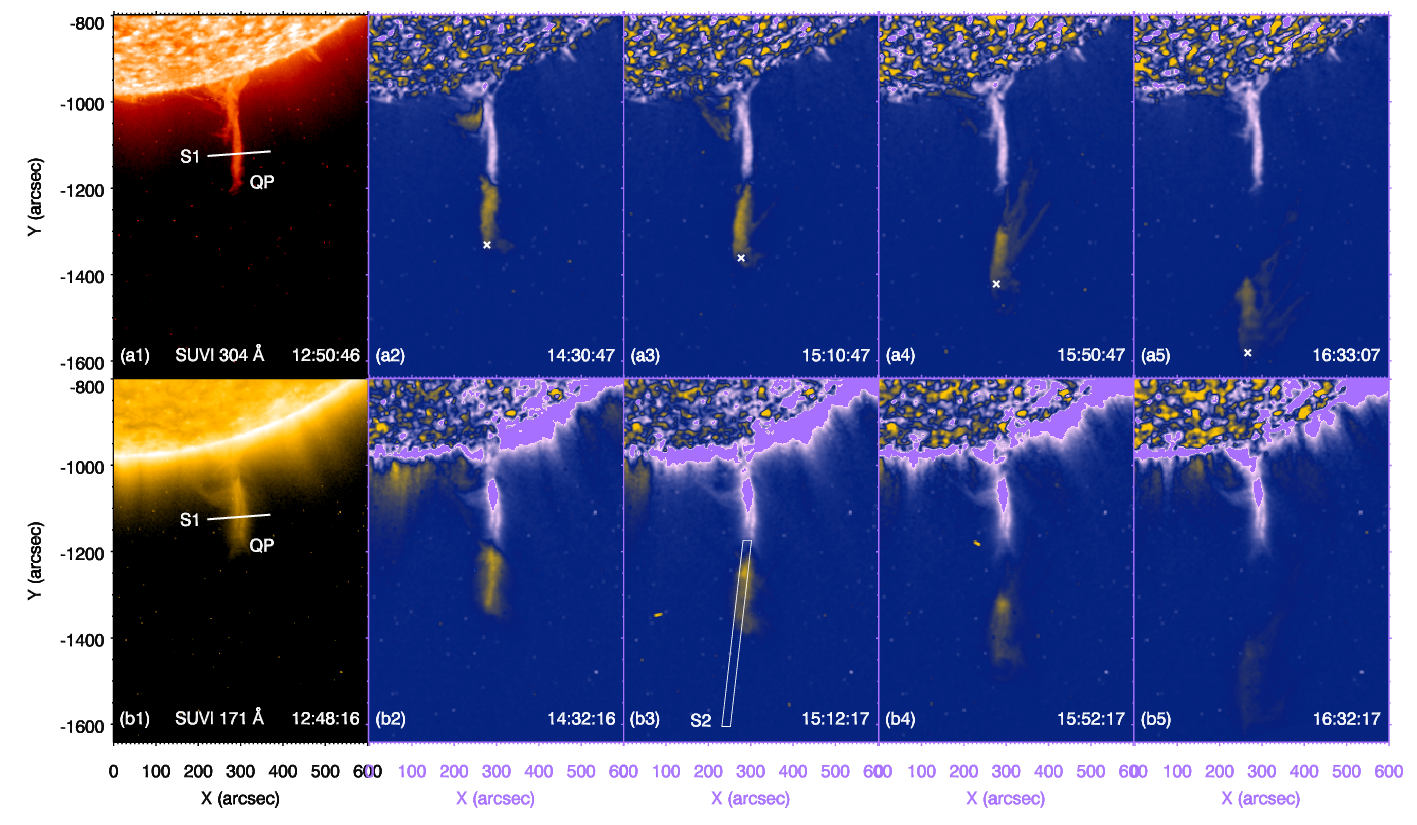}
    \caption{Top panels: SUVI 304 {\AA} original image at 12:50:46 UT and base-difference images during 14:30$-$16:33 UT.
    The white crosses mark the leading edges of QP.
    Bottom panels: SUVI 171 {\AA} original image at 12:48:16 UT and base-difference images during 14:32$-$16:32 UT.
    The slice (S1) in left panels is used to study the transverse oscillation of QP. The wide slice (S2) in panel (b3) is used to study the rising motion of QP.
    An online animation of the 304 and 171 {\AA} base-difference images is available. The $\sim$6 s animation covers from 12:50 UT to 16:50 UT.}
    \label{fig4}
\end{figure}

The left panels of Figure~\ref{fig4} show the QP with a position angle (PA) of $\sim$196$^{\circ}$, 
which is observed by SUVI 304 and 171 {\AA} around 12:49 UT.
The width and height of QP in 304 {\AA} are $\sim$17.4 Mm and $\sim$188 Mm, respectively. Therefore, the aspect ratio of QP is $\sim$11.

\begin{figure}
\centering
	\includegraphics[width=0.8\columnwidth]{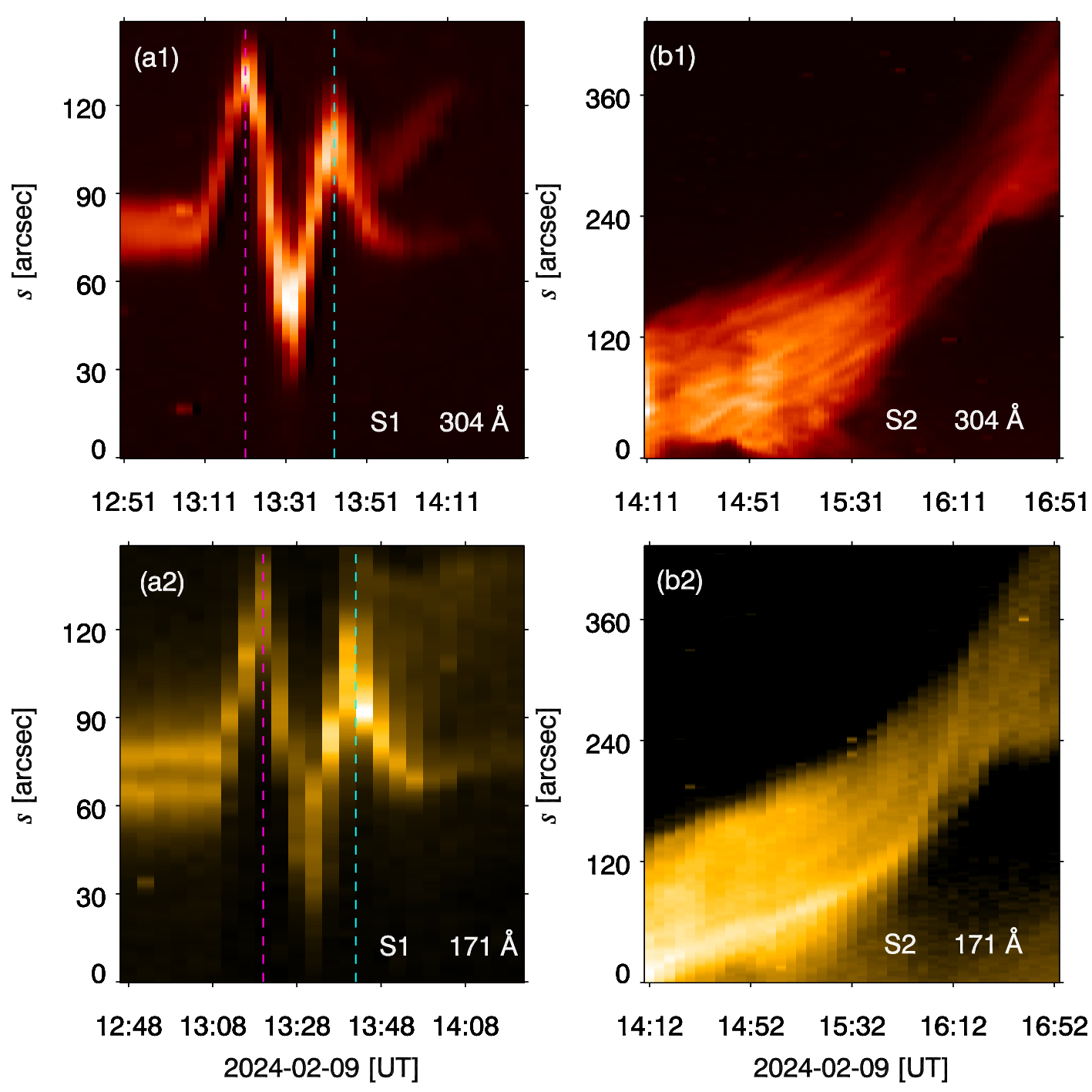}
    \caption{Time-distance diagrams of S1 (left panels) and S2 (right panels) in SUVI 304 {\AA} (top panels) and 171 {\AA} (bottom panels).
    The magenta and cyan dashed line indicate peak times of the transverse oscillation.}
    \label{fig5}
\end{figure}

The strong impact of WF1 excites a transverse oscillation of QP.
In the left panels of Figure~\ref{fig4}, a straight slice (S1) perpendicular to the prominence is selected to study the oscillation.
Time-distance diagrams of S1 in 304 and 171 {\AA} are displayed in the left panels of Figure~\ref{fig5}.
The magenta and cyan dashed line indicate peak times ($\sim$13:21 UT and $\sim$13:43 UT) of the transverse oscillation.
It is clear that QP swings back and forth for two cycles with an initial amplitude and a period of $\sim$27 Mm and $\sim$25 minutes.
It is worth mentioning that variations of the initial amplitudes, velocities, and periods with the prominence height are demonstrated in Fig. 10 of Paper I.

\subsection{Sympathetic eruption of QP} \label{symp}
The transverse oscillation lasts for up to $\sim$60 minutes till 14:09 UT. 
Then, QP breaks into two pieces. The top part lifts off to form CME2, while the lower part remains and collapses gradually.
Base-difference images in SUVI 304 and 171 {\AA} during 14:30$-$16:33 UT are displayed in panels (a2)-(a5) and panels (b2)-(b5) of Figure~\ref{fig4} 
(see also the online movie anim1.mp4 available as Supplementary material).
It is obvious that as QP rises up, a fraction of plasmas are falling back to the solar surface. The leading edges of QP in 304 {\AA} are marked with white crosses.
In Figure~\ref{fig1}(b), the magenta diamonds denote the heliocentric distances of the prominence leading edges, 
which increase from $\sim$1.40 $R_{\odot}$ to $\sim$1.68 $R_{\odot}$ during 14:30$-$16:50 UT.

\begin{figure}
\centering
	\includegraphics[width=0.8\columnwidth]{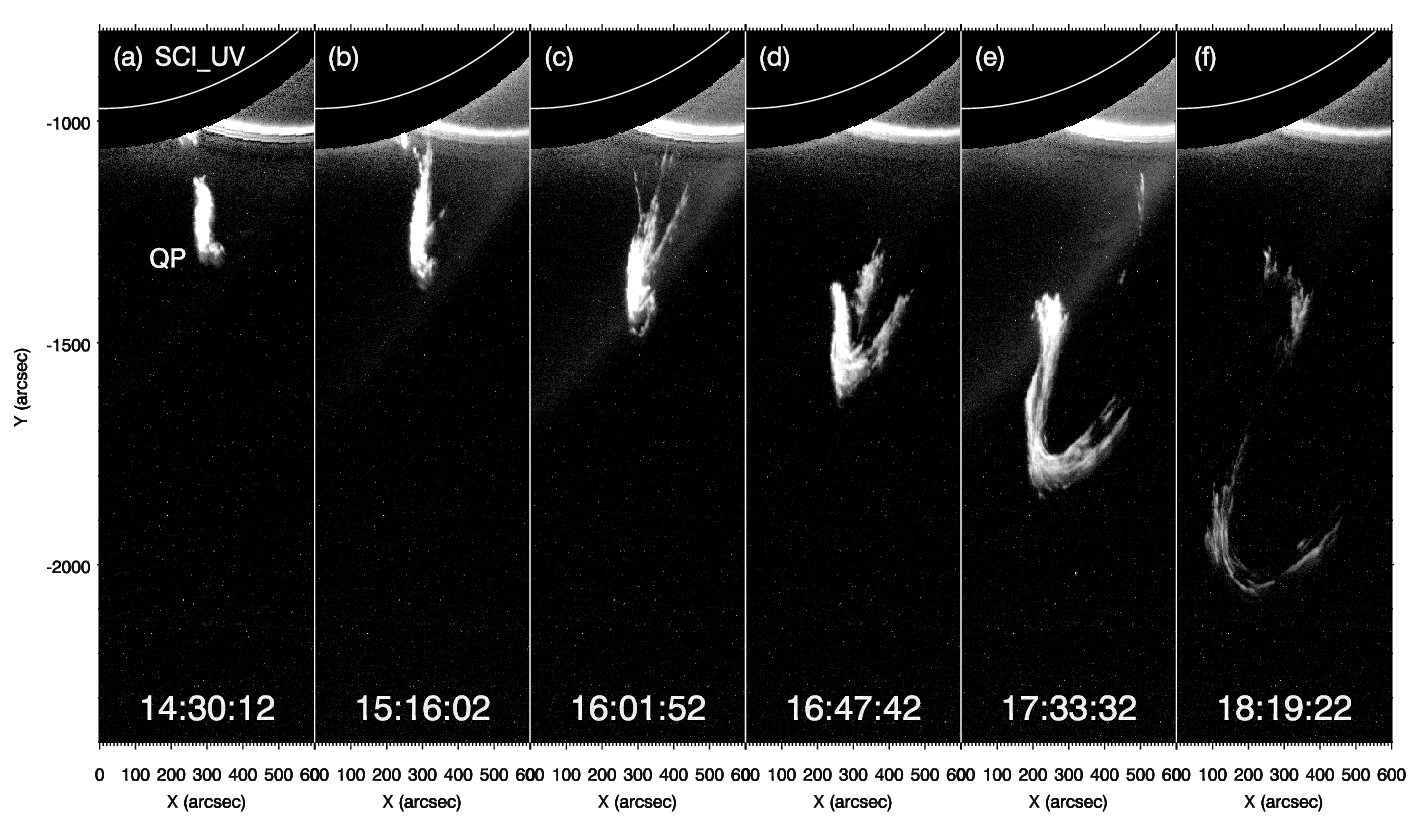}
    \caption{Rising QP observed by SCI\_UV during 14:30$-$18:20 UT.
    An online animation of the SCI\_UV images is available. The $\sim$13 s animation covers from 14:30 UT to 18:30 UT.}
    \label{fig6}
\end{figure}

In Figure~\ref{fig4}(b3), a wide slice (S2) is selected along the propagation direction of QP. 
It is evident that QP propagates non-radially with an inclination angle of $\sim$23$^{\circ}$ with the local vertical \citep{zqm22}.
Time-slice diagrams of S2 in 304 and 171 {\AA} are displayed in the right panels of Figure~\ref{fig5}. It is seen that QP is accelerating during the eruption.
As is shown in Table~\ref{tab-1}, SUVI has a FOV of $\sim$1.6 $R_{\odot}$. The erupting QP becomes blurred and fades out after 16:50 UT in SUVI images.
Fortunately, it is captured by the ASO-S/SCI\_UV coronagraph with a much larger FOV of $\sim$2.5 $R_{\odot}$ and a much higher time cadence.
Figure~\ref{fig6} shows QP observed by SCI\_UV during 14:30$-$18:20 UT (see also the online movie anim2.mp4 available as Supplementary material).
The shape of prominence changes from a column to a hook as it ascends.
Likewise, temporal evolution of the heliocentric distances of QP in the FOV of SCI\_UV are drawn with brown pluses in Figure~\ref{fig1}(b) and excellently fitted with a quadratic function:
\begin{equation} \label{eqn-1}
  \frac{h_{\mathrm{QP}}(t-t_0)}{R_{\odot}}=1.36+1.77\times10^{-5}(t-t_0)+2.80\times10^{-9}(t-t_0)^2,
\end{equation}
where $t_0$ is set to be 14:30:12 UT. $h_{QP}$ increases slowly from $\sim$1.37 $R_{\odot}$ and accelerates to $\sim$2.18 $R_{\odot}$ at 18:29:27 UT.
The apparent speed of QP is $v_{\mathrm{QP}}(t-t_0)=12.3+3.9\times10^{-3}(t-t_0)$ km s$^{-1}$.
Hence, $v_{\mathrm{QP}}$ increases slowly from 12.3 km s$^{-1}$ at 14:30 UT to 68.5 km s$^{-1}$ at 18:30 UT.

\begin{figure}
\centering
	\includegraphics[width=0.9\columnwidth]{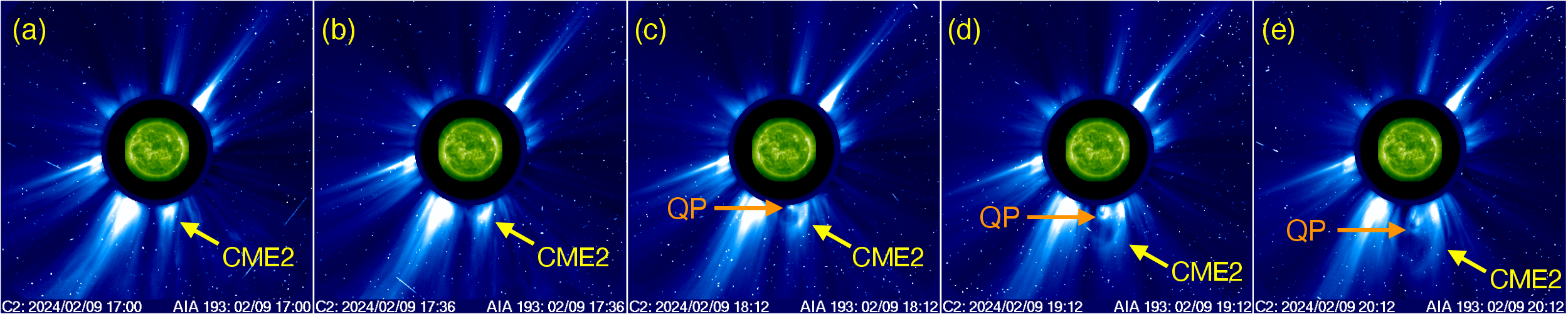}
    \caption{Original images observed by LASCO-C2 during 17:00$-$20:12 UT. 
    The rising QP and bright front of CME2 are pointed by orange and yellow arrows, respectively.}
    \label{fig7}
\end{figure}

Figure~\ref{fig7} shows five original images observed by LASCO-C2 during 17:00$-$20:12 UT. 
CME2 presents a typical three-part structure \citep{ill85,song22}.
The bright front of CME2, which is pointed by the yellow arrow, turns up at 17:00 UT and propagates in the south direction with a PA of 190$^{\circ}$ (panels (a)-(b)).
At 18:12 UT, the bright core of CME2 appears. Considering that CME2 is driven by the erupting QP, QP is exactly the core of CME2, which is pointed by an orange arrow (panel (c)).
Afterwards, QP and bright front of CME2 move outward together (panels (d)-(e)).
In Figure~\ref{fig1}(b), temporal evolution of the heliocentric distances of QP in the FOV of LASCO-C2 is plotted with blue circles.
It is obvious that QP evolves coherently in the FOVs of SCI\_UV and LASCO-C2. 
The slight difference in height during 18:12$-$18:30 UT is probably due to the different wavelengths.
QP is observed in UV (121.6$\pm$10 nm) by SCI\_UV and in WL by LASCO-C2, respectively.
The advantages of passband and large FOV of SCI\_UV enable us to track erupting prominences completely from their early phases up to $\sim$2.5 $R_{\odot}$.
In Figure~\ref{fig1}(b), temporal evolution of the heliocentric distances of the bright front is plotted with orange circles.
Similarly, the trajectory is satisfactorily fitted with a quadratic function:
\begin{equation} \label{eqn-2}
  \frac{h_{\mathrm{CME2}}(t-t_2)}{R_{\odot}}=2.60+1.73\times10^{-4}(t-t_2)+4.70\times10^{-9}(t-t_2)^2,
\end{equation}
where $t_2$ is set to be 17:00:00 UT. 
Likewise, the apparent speed of CME2 is $v_{\mathrm{CME2}}(t-t_2)=120.4+6.5\times10^{-3}(t-t_2)$ km s$^{-1}$.
Therefore, $v_{\mathrm{CME2}}$ increases gradually from 120.4 km s$^{-1}$ at 17:00 UT to 277.2 km s$^{-1}$ at 23:42 UT.

\begin{figure}
\centering
	\includegraphics[width=0.5\columnwidth]{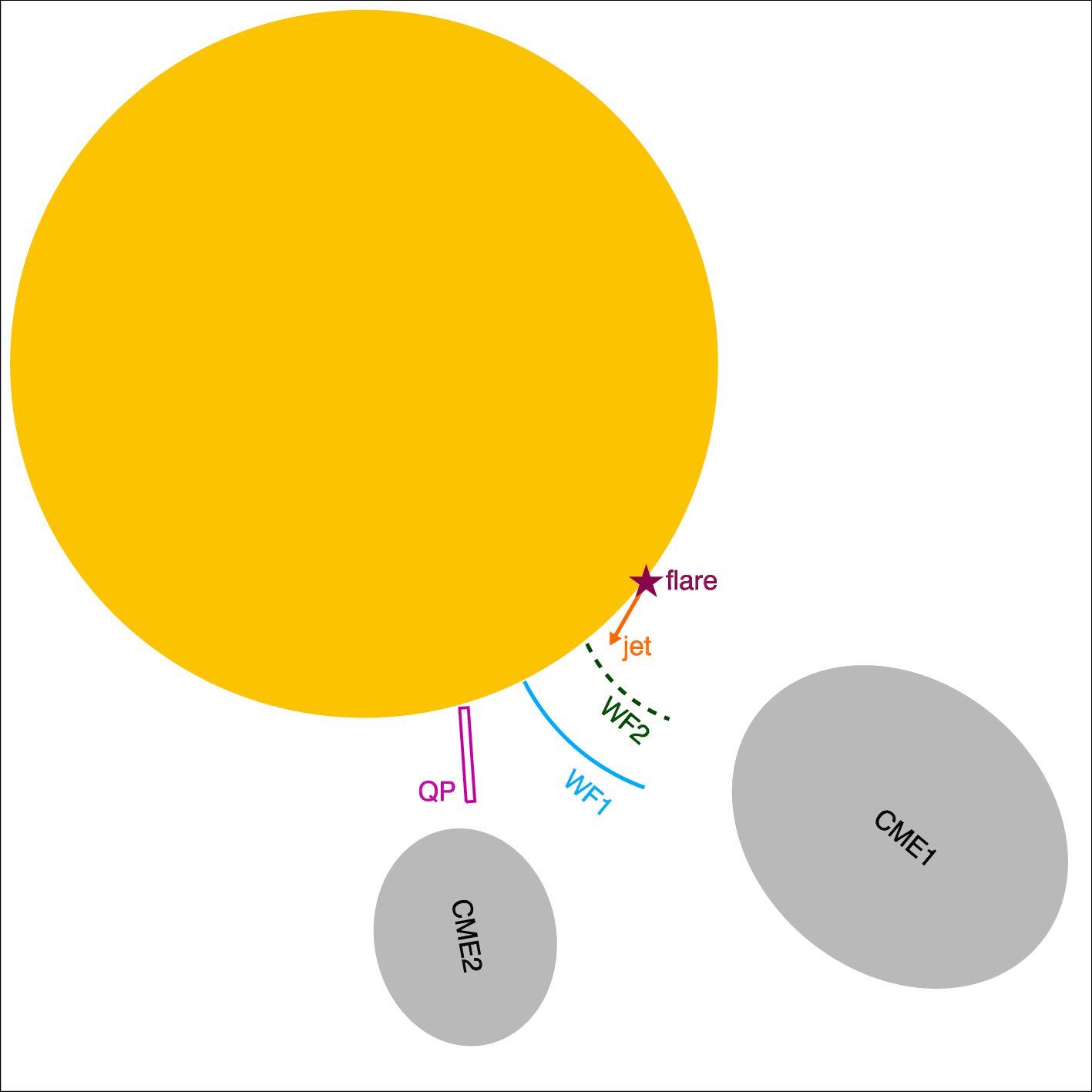}
    \caption{A schematic cartoon to illustrate the whole events on 2024 February 9.}
    \label{fig8}
\end{figure}

The whole events, including the X3.4 flare, CME1, coronal jet, EUV wave fronts (WF1 and WF2), QP, and CME2, are illustrated in a schematic cartoon in Figure~\ref{fig8}.
It is clear that the EUV waves, especially WF1, play an essential role in linking the primary and sympathetic eruptions.

\section{Discussion} \label{dis}
\citet{kol16} proposed a model of the global transverse oscillations and stability of quiescent prominences, which are supported by magnetic dips.
The periods of transverse oscillations with small amplitudes are derived analytically.
Considering that LAOs are frequently observed \citep{her11,shen14a,shen14b,dai23}, \citet{kol18} investigated the effects of finite amplitudes on the transverse prominence oscillations.
It is revealed that a metastable equilibrium of the prominence exists. The prominence is stable for small-amplitude displacements.
However, it gets unstable in the horizontal direction when the amplitude is large enough to exceed a threshold value.
In our study of the transverse oscillation of QP, the maximal displacement amplitude reaches $\sim$34 Mm and the maximal velocity amplitude reaches $\sim$143 km s$^{-1}$, 
which is close to the acoustic speed of the corona ($T\sim1.5$ MK) (see Fig. 10 in Paper I).
After the oscillation, QP starts to rise up and evolves into the bright core of CME2 (see Figure~\ref{fig7}).
Therefore, both the transverse oscillation and subsequent eruption of QP might be explained by their analytical model.
The large amplitudes of QP may exceed a threshold so that the prominence loses equilibrium and erupts, 
which provides a new evidence for the conclusion that large-amplitude transverse oscillation is one of precursors for prominence eruptions \citep{chen08}.

Similar metastable equilibrium may exist for a longitudinally oscillating prominence as well.
Using one-dimensional (1D) hydrodynamic numerical simulations with the MPI-AMRVAC code \citep{kep23},
\citet{zqm13} carried out a parameter survey of longitudinal prominence oscillations along magnetic dips.
It is found that the prominence is limited in the dips when the initial amplitudes are not so large.
Nevertheless, a fraction of the prominence reaches and overshoots the shoulders of dips, 
leading to mass drainage along the legs of the prominence and a possible eruption when the initial amplitude is large enough (see their Fig. 8).
\citet{fan20} performed three-dimensional (3D) magnetohydrodynamic (MHD) simulations of large-amplitude, longitudinal oscillations of a prominence supported by a twisted coronal flux rope.
It is revealed that the oscillations are quickly attenuated after a few cycles, which are followed by significant mass drainage and eventual eruption of the prominence.
From this point of view, LAOs of prominences, including transverse and longitudinal polarizations, are considered as a conceivable precursor of prominence eruptions.

\begin{table}
	\centering
	\caption{Comparison between the primary eruption and sympathetic eruption.}
	\label{tab-2}
	\begin{tabular}{ccccc}
		\hline
		Eruption & primary & sympathetic \\
		\hline
		Erupting structure & HC & QP \\
		Source region & AR 13575 & quiet region \\
		Start time & 12:49 UT & 14:10 UT \\
		Kinematics & slow rise and fast rise & constant acceleration \\
		Flare & X3.4 & $-$ \\
		CME & CME1 & CME2 \\
		CME type & impulsive & gradual \\
		CME speed [km s$^{-1}$] & 2782 & 277 \\
		EUV wave & WF1 & -- \\
		\hline
	\end{tabular}
\end{table}

\begin{figure}
\centering
	\includegraphics[width=0.6\columnwidth, angle=270]{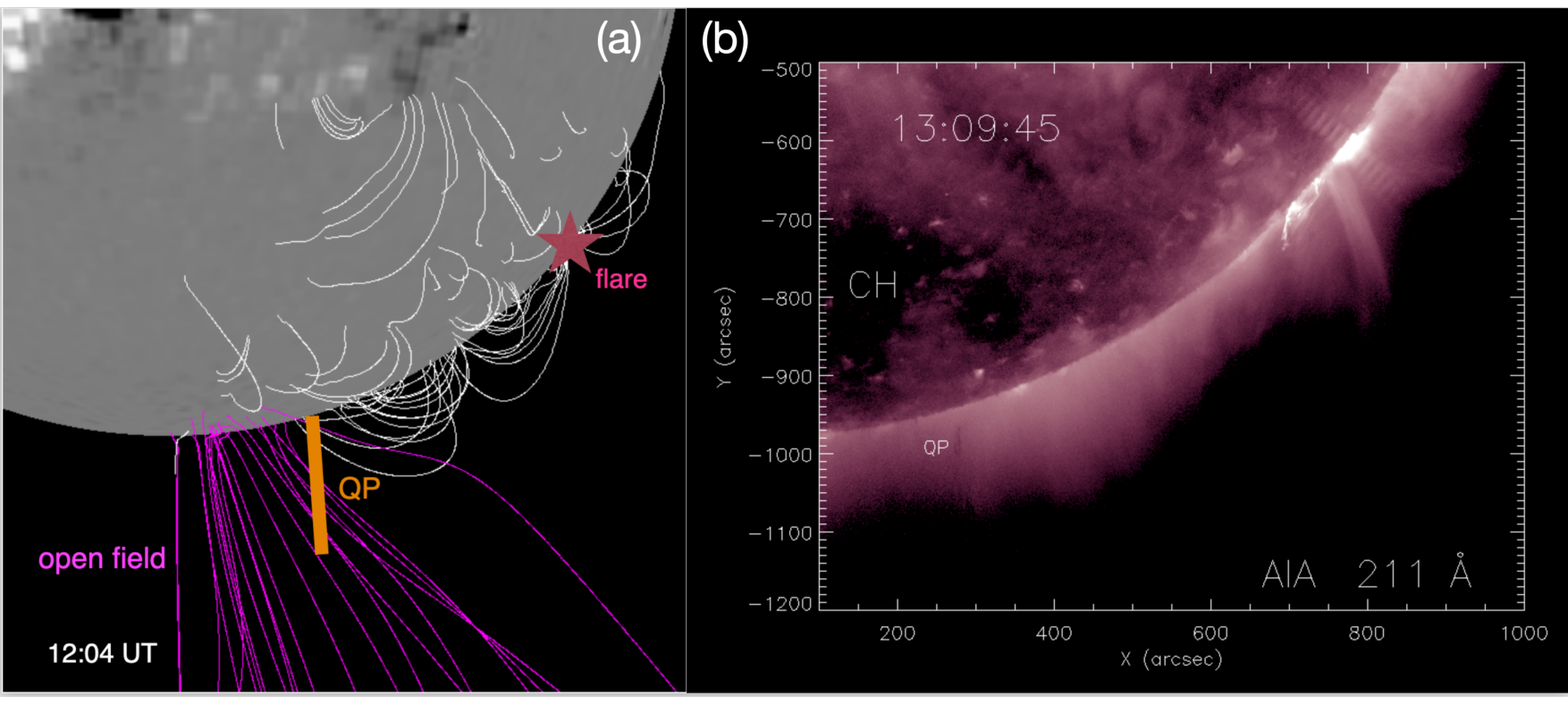}
    \caption{(a) Large-scale magnetic field lines obtained by the PFSS modeling at 12:04 UT on 2024 February 9.
    The open and closed fields are drawn with magenta and white lines, respectively.
    The flare and QP are denoted by a red star and an orange stick.
    (b) AIA 211 {\AA} image at 13:09:45 UT, showing the dark coronal hole (CH) and QP close to the south polar region.}
    \label{fig9}
\end{figure}

After developing a new technique for tracking CMEs in the FOV of LASCO, \citet{she99} divided their sample into two types, gradual CMEs and impulsive CMEs.
Impulsive CMEs are usually associated with flares and Moreton waves, while gradual CMEs are formed when prominences and their cavities rise up from below coronal streamers.
Besides, impulsive CMEs are generally faster and tend to decelerate during propagation, while gradual CMEs are relatively slower and tend to accelerate.
In Table~\ref{tab-2}, we compare the primary eruption and sympathetic eruption in detail.
The former results from a HC eruption at 12:49 UT from AR 13575, while the latter results from a prominence eruption at $\sim$14:10 UT from the quiet region, which is $\sim$80 minutes delayed.
The HC eruption is characterized by a slow rise and a fast rise phase \citep{zqm23,zqm24a}, while the prominence eruption presents a constant acceleration ($\sim$3.9 m s$^{-2}$).
The primary eruption generates an X3.4 class flare, an impulsive CME, and EUV wave (WF1) as described in Section~\ref{data}.
WF1 serves as a causal link between the consecutive eruptions. 
It is noted that the prominence does not erupt instantly after the arrival of WF1.
Instead, QP experiences a large-amplitude, transverse oscillation, which lasts for about one hour.
The prominence becomes unstable after the oscillation and erupts eventually.
The sympathetic eruption drives a gradual CME, whose leading edge shows a constant acceleration ($\sim$6.5 m s$^{-2}$).
The final speed of CME1 is almost 10 times higher than that of CME2.
In brief, the primary and sympathetic eruptions exhibit remarkably different properties.
\citet{zhang04} explored the kinematic properties of three CMEs associated with flares. 
It is found that there is close correlation both between the CME velocity and the SXR flux of the flare and between the CME acceleration and derivative of the SXR flux.
In our study, CME2 undergoes continuous acceleration for nearly seven hours. However, it is not related to a flare underneath.
In Figure~\ref{fig9}, the left panel shows large-scale magnetic field lines obtained by the potential-field source surface (PFSS; \citealt{sch69,sch03}) modeling at 12:04 UT.
The open and closed fields are indicated by magenta and white lines, respectively. The right panel shows the AIA 211 {\AA} image at 13:09:45 UT. 
Close to the south polar region, the coronal hole (CH) is fainter than the surrounding quiet region and is associated with the footpoints of open field lines in the left panel.
Since QP is also close to the south polar region, the slow CME is probably pushed by the fast solar wind during its propagation \citep{has99,tu05}.

Using He\,{\sc ii} 304 {\AA} observations with the Full Sun Imager (FSI) of the Extreme Ultraviolet Imager \citep[EUI;][]{roc20} onboard the Solar Oribiter \citep[SolO;][]{mu20} mission,
\citet{mie22} reported a fast prominence eruption up to $>$6 $R_{\odot}$ on 2022 February 15$-$16.
The velocities of the leading edges of prominence and associated CME are $\sim$1700 and $\sim$2200 km s$^{-1}$, respectively.
Using combined observations with the EUI/FSI in 174 {\AA} and the SolO/Metis \citep{ant20} in visible light (VL) and UV (Ly$\alpha$),
\citet{bem22} studied a CME followed by a prominence eruption and a long current sheet on 2021 February 12.
The prominence was tracked by Metis from $\sim$3 to $\sim$5.4 $R_{\odot}$, while the early phase of the prominence was not observed due to the occulting disk.
In our case, the prominence is completely tracked by ASO-S/SCI\_UV from $\sim$1.1 to $\sim$2.2 $R_{\odot}$ in Ly$\alpha$ wavelength, which is crucial to study the kinematics.

\section{Summary} \label{sum}
In this paper, we perform a follow-up investigation of the solar eruption originating from AR 13575 on 2024 February 9. The main results are summarized as follows:
\begin{enumerate}
\item The primary eruption of a HC results in an X3.4 class flare, a full-halo CME (CME1), and an EUV wave (WF1).
Interaction between WF1 and QP leads to a large-amplitude, transverse oscillation of QP, which have been described in our previous paper (Paper I).
\item After the transverse oscillation, QP becomes unstable and lifts off.
The rising motion of the prominence is clearly detected and tracked by GOES-16/SUVI until $\sim$1.68 $R_{\odot}$ 
and by LST/SCI\_UV onboard the ASO-S spacecraft until $\sim$2.2 $R_{\odot}$.
The velocity increases linearly from 12.3 to 68.5 km s$^{-1}$ at 18:30 UT.
The sympathetic eruption of QP drives the second CME (CME2) with a typical three-part structure. 
The bright core comes from the eruptive prominence, which could be tracked further until $\sim$3.3 $R_{\odot}$ by SOHO/LASCO-C2.
The leading edge of CME2 accelerates continuously from $\sim$120 to $\sim$277 km s$^{-1}$.
\item The EUV wave serves as a causal link between the primary and sympathetic eruptions.
The advantageous cadence and FOV of ASO-S/SCI\_UV provide a great opportunity to track an erupting prominence from $\sim$1.1 to $\sim$2.2 $R_{\odot}$ in Ly$\alpha$ wavelength.
\end{enumerate}

\begin{acknowledgements}
The authors appreciate the reviewer for valuable suggestions to improve the quality of this article.
We also thank Prof. Hui Li in Purple Mountain Observatory, Prof. Pengfei Chen in Nanjing University, and Prof. Xiaoli Yan in Yunnan Astronomical Observatories for helpful discussions.
SDO is a mission of NASA\rq{}s Living With a Star Program. AIA data are courtesy of the NASA/SDO science teams.
This CME catalog is generated and maintained at the CDAW Data Center by NASA and The Catholic University of America in cooperation with the Naval Research Laboratory. 
SOHO is a project of international cooperation between ESA and NASA.
The ASO-S is supported by the Strategic Priority Research Program on Space Science, Chinese Academy of Sciences.
This work is supported by the National Key R\&D Program of China 2022YFF0503003 (2022YFF0503000), 2021YFA1600500 (2021YFA1600502), 
the Strategic Priority Research Program of the Chinese Academy of Sciences, Grant No. XDB0560000,
the NSFC under the grant numbers 12373065, 12203102, 12403064, 12403068, Natural Science Foundation of Jiangsu Province (BK20231510, BK20241707), 
the Project Supported by the Specialized Research Fund for State Key Laboratories,
and Yunnan Key Laboratory of Solar Physics and Space Science under the grant number YNSPCC202206.
\end{acknowledgements}

\label{lastpage}

\end{document}